%% file: dm_space_distortions.tex
\documentclass[twocolumn,nofootinbib,prl,floatfix]{revtex4-1}
\usepackage{graphicx}
\usepackage{amsmath}
\usepackage{amssymb}
\usepackage{bm}
\usepackage{color}
\usepackage{verbatim}
\usepackage{times}
\usepackage{afterpage}
\usepackage{epstopdf}
\usepackage{url}
\usepackage{hyperref}

\newcommand{\ud}{\,d}

\newcommand{\blue}{}

\makeatletter
\def\ignorespacesandimplicitepars{%
  \begingroup
  \catcode13=10
  \@ifnextchar\relax
    {\endgroup}%
    {\endgroup}%
}

\renewcommand{\section}[1]{\emph{#1.}---\ignorespacesandimplicitepars}
\renewcommand{\subsection}[1]{}

\input{AAS_macros.tex}

\makeatother

\begin{document}

\title{Dispersion Distance and the Matter Distribution of the Universe in
Dispersion Space}

\author{Kiyoshi Wesley Masui}
 \affiliation{Department of Physics and Astronomy, University of British 
Columbia, Vancouver, British Columbia, V6T 1Z1, Canada}
 \affiliation{Canadian Institute for Advanced Research, CIFAR Program in
Cosmology and Gravity, Toronto, Ontario, M5G 1Z8, Canada}

\author{Kris Sigurdson}
 \affiliation{Department of Physics and Astronomy, University of British 
Columbia, Vancouver, British Columbia, V6T 1Z1, Canada}

\begin{abstract}

    We propose that ``standard pings'', brief broadband radio impulses, can be
    used to study the three-dimensional clustering of matter in the Universe
    even in the absence of redshift information.  The dispersion of radio waves
    as they travel through the intervening plasma can,
    like redshift, be used as a
    cosmological distance measure.  Because of inhomogeneities in the electron
    density along the line of sight, dispersion is an imperfect proxy for
    radial distance and we show that this leads to calculable dispersion-space
    distortions in the apparent clustering of sources. Fast radio bursts (FRBs)
    are a new class of radio transients that are the prototypical standard
    ping and, due to their high observed dispersion,
    have been interpreted as originating at cosmological distances.  The rate
    of fast radio bursts has been estimated to be several thousand over the
    whole sky per day and, if cosmological, the sources of these events should
    trace the large-scale structure of the Universe.  We calculate the
    dispersion-space power spectra for a simple model where electrons and FRBs
    are biased tracers of the large-scale structure of the Universe and we show
    that the clustering signal could be measured using as few as
    $10\,000$ events. Such a survey is in line with what may be achieved with
    upcoming wide-field radio telescopes.
\end{abstract}

\maketitle

\section{Introduction}

The clustering of matter on large scales has been
heralded as the next great probe of the Universe after the cosmic microwave
background (CMB). The large-scale structure, in principle, contains far more
information than the CMB because it can be studied in three dimensions.
Traditionally, the redshift of spectral lines of ``standard atoms'', 
caused by the Hubble flow, has
been employed to measure radial distance and provide the third dimension.
Such redshift
surveys have now aggregated the positions of millions of galaxies into three
dimensional density maps, exposing a rich structure of clusters, filaments,
walls, and voids.

However, redshift is not the only measure of radial distance. Standard
candles and standard rulers can be used to estimate distance using brightness
and angular size, respectively.  In the same way a ``standard ping'', a brief
broadband radio impulse, can be used to measure dispersion through the
characteristic wavelength-squared delay of electromagnetic signals. The dispersion is
proportional to the amount of plasma along the
line of sight and can thus be used to estimate radial distance.
This technique, routinely used for
pulsars in the Galaxy, can be used to define a \emph{dispersion distance} to
cosmological sources.

Dispersion is readily measured from fast radio bursts (FRBs),
a new class of millisecond duration radio transients first reported by
\citet{2007Sci...318..777L}, which are the archetype of a standard ping.
Ten such bursts have been reported to date
\citep{2012MNRAS.425L..71K,2013Sci...341...53T, 2014ApJ...790..101S,
2014ApJ...792...19B, 2015MNRAS.447..246P,2015ApJ...799L...5R}, and the total
event rate has been estimated to be $3.3^{+5}_{-2.5} \times 10^3$ per sky per
day \cite{2015arXiv150500834R}.
Dispersion is characterized by a line-of-site
integral of the free electron density known the dispersion measure (DM). FRB
events are observed to have dispersion measures of order $\sim\!\!1000\,{\rm pc}
/ {\rm cm}^3$, greatly in excess of the galactic expectation of order
$\sim\!\!100\,{\rm pc} / {\rm cm}^3$ (depending on galactic latitude)
\cite{2015MNRAS.451.4277D}.  They thus appear to be of extragalactic origin,
with their dispersion produced either by the intergalactic medium,
associated with structure along line of sight \citep{2013Sci...341...53T},
or the environment of the
source \citep{2014ApJ...785L..26L,2015ApJ...807..179P,2015arXiv150505535C}.  If
the dispersion is dominated by electrons along the line of sight, then the
average dispersion--distance relation can be modeled
\citep{2014ApJ...797...71Z,2004MNRAS.348..999I, 2003ApJ...598L..79I} and the
sources appear to be at cosmological distances of order gigaparsecs.
Such cosmological distances permit the study of the clustering of FRB sources in
three dimensions, with an expectation that these events trace the large-scale
structure of the Universe on linear scales. We note that while our discussion
here is framed in terms of FRBs, other bright radio transients that fluctuate
on time scales shorter than the typical differential lag of $\sim 1s/100 \,{\rm
MHz} $ might also be useful as standard pings.

Neither redshift nor dispersion are perfect proxies for radial distance.
Redshifts are systematically biased by the peculiar velocities of objects
relative to the Hubble flow. This leads to additional apparent clustering
in redshift
space which was first calculated by \citet{1987MNRAS.227....1K}
using linear theory.
Similarly, as we show here, distance estimates from dispersion will be biased by
inhomogeneities in the electron density \citep{2014ApJ...780L..33M},
again leading to additional apparent
clustering of sources in dispersion space.  In this Letter we calculate
these dispersion-space distortions and consider the detectability of the signal by
upcoming surveys.

\section{Clustering in dispersion space}

The dispersion measure of a signal observed in some angular direction $\hat{n}$ and
originating from comoving radial distance $\chi$ is
\begin{equation}
    {\rm DM}(\hat{n}, \chi) = \int_0^{\chi} \ud\chi' a(\chi)^2
        n_e(\hat{n}\chi', \chi').
\end{equation}
Here
$n_e(\vec{x}, \chi)$ is the free electron density as a function of location and
conformal time. Note that we use $\chi$ as our radial distance and time
coordinate, as opposed to redshift. We model the cosmological 
electron density as containing a homogeneous part
and perturbations,
$n_e(\vec x, \chi) = \bar{n}_e(\chi) \left[1 + \delta_e(\vec x, \chi)\right]$.
The dispersion
measure is thus
\begin{equation}
    {\rm DM}(\hat n, \chi) = \int_0^\chi \ud\chi' a(\chi')^2 \bar{n}_e(\chi')
       \left[1 + \delta_e(\hat{n}\chi',\chi')\right].
\end{equation}

\emph{Dispersion space} is the three-dimensional coordinates,
$\vec x_s$, inferred from the dispersion measure assuming that the electrons are
homogeneous. This only affects the radial coordinate such that $\vec x_s = \hat
n \chi_s$, with $\chi_s$ defined by the equation
\begin{equation}
    {\rm DM}(\chi_s) = \int_0^{\chi_s} \ud\chi' a(\chi')^2 \bar{n}_e(\chi').
\end{equation}

Combining the above equations (keeping only terms first order in the density
perturbations), we find that
\begin{equation}
\frac{\ud \chi_s}{\ud \chi} = 1 + \delta_e(\hat n \chi, \chi),
\end{equation}
and thus
\begin{equation}
\chi_s - \chi = \int_0^\chi \ud \chi' \delta_e(\hat n \chi').
\end{equation}

We wish to relate the density of a tracer, $f$, measured in dispersion space to its
density in real space. We follow the derivation in \citet{1987MNRAS.227....1K}
of the
redshift-space distortions. Start by noting that the total number of tracers
in a volume element is the same in both spaces:
\begin{equation}
n_{fs}(\vec x_s) \ud^3\vec x_s = n_{f}(\vec x) \ud^3\vec x.
\end{equation}
We split the density into a homogeneous part plus perturbations,
\begin{equation}
\label{e:density}
    \bar{n}_{fs}(\chi_s)\left[ 1 + \delta_{fs}(\vec x_s)\right] \ud^3\vec x_s
    = \bar{n}_{f}(\chi)\left[ 1 + \delta_f(\vec x)\right] \ud^3\vec x.
\end{equation}
Averaged over the sky, $\langle \chi_s \rangle = \chi$, and thus the background
density should
be the same in dispersion space as in real space,
\begin{equation}
\bar{n}_{fs}(\chi) = \bar{n}_f(\chi).
\end{equation}
Therefore,
\begin{align}
\bar{n}_{fs}(\chi_s) 
    &= \bar{n}_{f}(\chi) + (\chi_s - \chi)\frac{\ud \bar{n}_f}{\ud \chi}\\
    &= \bar{n}_{f}(\chi)
       + \frac{\ud \bar{n}_f}{\ud \chi}\int_0^\chi \ud \chi' \delta_e(\hat n
       \chi').
       \label{e:nfs}
\end{align}

The Jacobian in spherical coordinates is
\begin{align}
\left| \frac{\ud^3\vec x}{\ud^3\vec x_s} \right|
    &= \frac{\ud \chi}{\ud \chi_s}\frac{\chi^2}{\chi_s^2}\\
    &\approx 1 - \delta_e
        - \frac{2}{\chi}\int_0^\chi \ud \chi' \delta_e(\hat n \chi').
        \label{e:jac}
\end{align}

Substituting Eqs.~\ref{e:nfs}~and~\ref{e:jac} into Eq.~\ref{e:density},
we obtain
\begin{equation}
\label{e:delta_s}
    \delta_{fs} = \delta_f - \delta_e
    - \left(\frac{1}{\bar{n}_f}\frac{\ud \bar{n}_f}{\ud \chi}
    + \frac{2}{\chi} \right)
        \int_0^\chi \ud \chi' \delta_e(\hat n \chi').
\end{equation}
In the above equation, the $-\delta_e$ term is most analogous to the Kaiser
redshift-space distortions.  It is a dilution of tracers in dispersion space due to an
excess of electrons between the tracers. However we note that this term is isotropic
in contrast to the Kaiser term.  This is because any wave vector electron perturbation
causes dispersion-space distortions, whereas the radial velocities that cause
redshift-space distortions are only sourced by perturbations with radial wave
vectors.

The $\frac{1}{\bar{n}_f}\frac{\ud \bar{n}_f}{\ud \chi}$ term arises because the
misinterpretation of the radial distance causes the observed tracer density to
be compared to the wrong background density. The $\frac{2}{\chi}$ term is
caused by a misinterpretation of angular distances when the radial distance is
mismeasured.  In both cases, analogous terms are, in principle, present in 
redshift space but are
negligible. Because radial velocities are only sourced by modes with a large
radial wave vector, there is near perfect cancellation along the line of sight,
and thus there is very little net error in the radial distance.

For brevity in the following sections, we define the coefficient of the integral
term as
\begin{equation}
    \label{e:A}
    A(\chi) \equiv \frac{1}{\bar{n}_f}\frac{\ud \bar{n}_f}{\ud \chi}
    + \frac{2}{\chi}.
\end{equation}


Large-scale structure is usually studied through its two-point statistics, most
commonly the power spectrum, $P(k)$.
Unlike the redshift-space distortions, Eq.~\ref{e:delta_s} does not have a
simple form in harmonic space. The equation's third term couples harmonic
modes,
and thus the two-point statistics cannot be phrased as a simple power spectrum.
We will instead use $C^{ss}_\ell(\chi,\chi')$, which is
the cross-correlation angular power spectrum of the dispersion-space overdensity,
on shells at $\chi$ and $\chi'$:
\begin{align}
    &C^{ss}_\ell(\chi, \chi') =
    \int\ud\Omega\ud\Omega' Y_{\ell m}(\hat n) Y^*_{\ell m}(\hat n')
    \langle \delta_{fs}(\hat n \chi) \delta_{fs}(\hat n' \chi')
        \rangle.
\end{align}
The first two terms in Eq.~\ref{e:delta_s} are stationary. For stationary,
isotropic tracers $x$ and $y$, we have 
$\langle \delta_x(\vec k, \chi) \delta_y(\vec k', \chi) \rangle = (2\pi)^3
\delta^3(\vec k - \vec k') P_{xy}(k, \chi)$.  If, for a moment, we ignore
structure evolution, the angular cross-power spectrum of such tracers is
\begin{align}
    C^{xy}_\ell(\chi,\chi')
    &= \frac{2}{\pi}
\int_0^\infty\ud k k^2 j_\ell(k\chi) j_{\ell}(k\chi')P_{xy}(k).
\end{align}
In reality the power spectrum evolves on the order of a Hubble time.
The angular cross correlations will be very small unless $\chi$ and $\chi'$ are
within a few correlation lengths of one another, roughly a hundred
megaparsecs.  The evolution of the power spectrum is negligible over these time
differences, which leads to a straightforward way to include the evolution:
\begin{align}
    C^{xy}_\ell(\chi,\chi') \approx&
        \frac{2}{\pi}
        \int_0^\infty\ud k k^2
        j_\ell(k\chi) j_{\ell}(k\chi')
        P_{xy}(k,(\chi + \chi')/2),\nonumber \\
    &|\chi - \chi'| \ll 1/aH.
\end{align}

The third term in Eq.~\ref{e:delta_s} is not stationary but is an integral
over the stationary field $\delta_e$.
Define $\delta_d$ as
\begin{equation}
    \delta_d(\hat n \chi) \equiv \int_0^\chi \ud \chi' \delta_e(\hat n \chi').
\end{equation}
It is straightforward to show that
\begin{widetext}
\begin{equation}
C^{dd}_\ell(\chi,\chi')
    =
    \frac{2}{\pi}
    \int_0^\chi\ud\chi''
    \int_0^{\chi'}\ud\chi'''
    \int_0^\infty\ud k k^2 j_\ell(k\chi'') j_{\ell}(k\chi''')
    P_{ee}(k, (\chi''+\chi''')/2).
\end{equation}

Finally, $C^{ss}_\ell$ will contain cross terms between the stationary terms
and the integral terms.  These will have the form
\begin{equation}
C^{dx}_\ell(\chi,\chi')
    =
    \frac{2}{\pi}
    \int_0^\chi\ud\chi''
    \int_0^\infty\ud k k^2 
    j_\ell(k\chi') j_{\ell}(k\chi'')P_{ex}(k, (\chi' + \chi'')/2).
\end{equation}

Assembling all of these expressions with the proper coefficients, we have
\begin{align}
    \label{e:Clss}
C^{ss}_\ell(\chi,\chi') = &~
    \frac{2}{\pi}
    \int_0^\infty\ud k k^2
    j_\ell(k\chi) j_{\ell}(k\chi')
    P_{[ff + ee - 2ef]}(k, (\chi + \chi')/2)
    \nonumber\\
    & +
    \frac{2}{\pi}
    A(\chi) A(\chi')
    \int_0^\chi\ud\chi''
    \int_0^{\chi'}\ud\chi'''
    \int_0^\infty\ud k k^2 j_\ell(k\chi'') j_{\ell}(k\chi''')
    P_{ee}(k, (\chi''+\chi''')/2)
    \nonumber\\
    & +
    \frac{2}{\pi}
    A(\chi)
    \int_0^\chi\ud\chi''
    \int_0^\infty\ud k k^2 
    j_\ell(k\chi') j_{\ell}(k\chi'')
    P_{[ee - fe]}(k, (\chi' + \chi'')/2)
    \nonumber\\
    & +
    \frac{2}{\pi}
    A(\chi')
    \int_0^{\chi'}\ud\chi''
    \int_0^\infty\ud k k^2 
    j_\ell(k\chi) j_{\ell}(k\chi'')
    P_{[ee - fe]}(k, (\chi + \chi'')/2).
\end{align}
\end{widetext}
Here, expressions such as $P_{[ff + ee - 2ef]}$ are shorthand for $P_{ff} +
P_{ee} - 2P_{ef}$.

Equation~\ref{e:Clss} can be simplified substantially by adopting the small
angle and Limber approximations \citep{1953ApJ...117..134L,
1992ApJ...388..272K, 1998ApJ...498...26K}.  The small angle
approximation eliminates the $k$ integral over spherical Bessel functions, replacing it
with a Fourier transform, and is valid for $\ell \gg 1$. The Limber
approximation assumes that only modes with a small radial component of their wave
vector contribute to the radial integrals and is valid if the power spectra
evolve slowly compared to the correlation length \citep{2008PhRvD..78l3506L}
(which has already been
assumed). With these approximations, we have
\begin{align}
    &C^{ss}_\ell(\chi, \chi') \approx
    \nonumber \\ & \quad
    \frac{1}{\bar\chi^2}
        \int_{-\infty}^\infty\frac{\ud k_\parallel}{(2 \pi)} 
        e^{i k_\parallel (\chi - \chi')}
        P_{[ee + ff - 2ef]}
        (\sqrt{k_\parallel^2 + \nu^2/\bar\chi^2}, \bar\chi)
    \nonumber \\ & \quad +
    A(\chi) A(\chi')
    \int_0^{\chi_{\rm min}}\ud\chi''
    \frac{1}{\chi^{\prime\prime 2}}
    P_{ee}(\nu/\chi'',\chi'')
    \nonumber \\ & \quad +
    \frac{A(\chi_{\max})}{\chi_{\min}^2}
    P_{[ee - ef]}(\nu/\chi_{\min}, \chi_{\min}),
    \label{e:Clss_limber}
\end{align}
where $\nu\equiv\ell + 1/2$, $\chi_{\min} \equiv \min(\chi, \chi')$,
and $\chi_{\max} \equiv \max(\chi,
\chi')$ .  We have found that these approximations are accurate to within $3\%$ at
$\ell \gtrsim 10$
and use this form for the remainder of the Letter. In the above equation, we dub
the three terms the ``local'', ``integral'' and ``cross'' terms, respectively,
and we will
refer to them as such henceforth.

\section{Modelling and measurement}

The second term in Eq.~\ref{e:Clss_limber} is an integral that mixes
many spatial scales: small scales from smaller radial distance and larger
scales from greater distances. The mixing of scales complicates the
interpretation of measurements since structure formation on scales smaller
than $\sim10\,{\rm Mpc}/h$ is nonlinear and hard to model. A similar issue
exists in the field of weak gravitational lensing, where the formula for the
shear angular power spectrum has a similar form.  In lensing, a kernel in
the line-of-sight integral arising from geometric effects suppresses the
contribution from small distances, partially alleviating the issue.

However, the lack of a kernel simplifies tomography---the use of sources at
multiple redshifts to unmix spatial scales.
Lensing tomography is inexact because the kernel's shape depends on
the source redshift, thus reweighting the line-of-sight integral.
In contrast, the integrand in the dispersion-space integral
term is completely independent of the limit of integration, meaning
contributions from different parts of the line of sight (and thus different
spatial scales) can be separated exactly. We leave a study of the potential
of dispersion-space
tomography to future work.  To avoid our results being
sensitive to the hard-to-model small scales, we assume that the contribution to the
integral term from radial distances $\chi'' < 500\,{\rm Mpc}/h$ are well
measured by the correlations with these radial slices. We thus ignore
contributions to the integral term from below this distance and neglect the
information from these slices in our sensitivity measurements. The smallest
spatial scale that contributes to our plotted angular power spectra out to
$\ell=1000$ is thus $k=2\,h/{\rm Mpc}$.

\begin{figure}
    \includegraphics[scale=0.42]{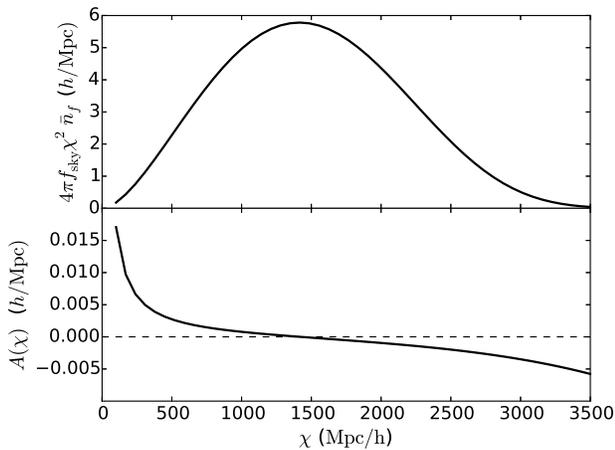}
    \caption{
        \label{f:n_f}
        (Top) Comoving density of sources in toy model, as described in
        text. Normalization is
        such that the survey has a total of $10\,000$ FRB events.
        (Bottom) Resulting coefficient $A(\chi)$ as given in
        Eq.~\ref{e:A}.
    }
\end{figure}

\begin{figure}
    \includegraphics[scale=0.42]{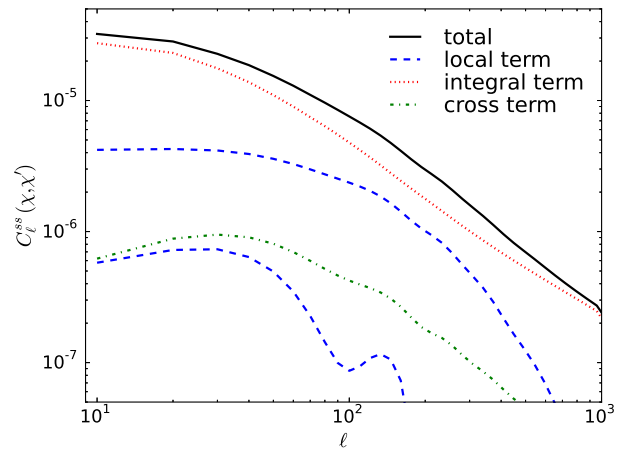}
    \caption{
        \label{f:terms}
        Terms in the dispersion-space cross-correlation angular power spectrum.
        All terms are evaluated at $\bar\chi=2000\,{\rm Mpc}/h$ and 
        $\chi - \chi' = 10\,{\rm Mpc}/h$, except the lowest most 
        local term curve, which is for $\chi - \chi'=50\,{\rm Mpc}/h$. 
        While
        the other two terms are highly insensitive to the separation of
        the radial slices, the local term drops rapidly with separation.
    }
\end{figure}

For plotting the dispersion-space power spectra,
we use a toy model for the electron and
FRB clustering that assumes both are biased tracers of the dark matter:
\emph{i.e.}~$P_{xy}(k)=b_x b_y P(k)$.  We
calculate the time dependant dark-matter power spectrum using {\tt CAMB}, using
the integrated {\tt HALOFIT} to approximate the nonlinear evolution. We assume
that the electron bias is $b_e=1$, which should be true on large scales due to
the approximate conservation of free electrons. The bias of FRB sources is
unconstrained by data and we choose it to be $b_f=1.3$\blue{, roughly
the value for star forming galaxies at the relevant
redshifts \citep{2011MNRAS.415.2876B} and providing reasonable contrast to the
electrons}.
The coefficient
$A(\chi)$ depends on the density of detected FRB sources, $\bar n_f(\chi)$.
We use a simple model where all events above a fixed flux are detected,
the intrinsic luminosity function has the form from
\citet{1976ApJ...203..297S} with index $-1$, and events at the cutoff of the
luminosity function and at
a radial distance of $\chi=2350\,{\rm Mpc}/h$ are at the flux detection threshold.
This source density is shown in Fig.~\ref{f:n_f} with the resulting dispersion-space
power spectrum in Fig.~\ref{f:terms}.

\blue{
The amplitudes of the local and cross terms depend on the difference
between the electron bias and the highly uncertain FRB bias,
and the sign of the cross term depends on
which of these is larger. However, since the
integral term dominates for most radial distance pairs, our final sensitivity
estimates are largely unaffected by this uncertainty (as shown below). In a
high-precision survey, the sensitivity of the cross and local terms to the
difference in the biases may help break measurement degeneracies, while sign
changes of terms as a function of radial distance are a signature of
bias evolution.
}

In a real survey, measurement of the dispersion-space clustering
will be complicated by extra contributions to the dispersion.
First, all signals
must pass through the Milky Way's galactic disk, halo, and local environment.
However, the resulting dispersion is only a
single function of $\hat n$ which presumably can be well
measured. As such, it should be
possible to subtract the local
contribution, as was done in
\citet{2015MNRAS.451.4277D}, and this is unlikely to be a limiting obstacle.

Perhaps more concerning is the dispersion from the source's immediate environment
as well as the halo in which it resides. These electrons will presumably be
clustered around the source on scales far smaller than the survey
resolution and are thus not accounted for in the
$\delta_e$--$\delta_f$
correlations included in the above formalism.  This will have a mean
contribution, which is common to all sources and which may be epoch dependant, 
that will
modify the mean dispersion--distance relation. This will need to be calibrated in some
way or else fit with nuisance parameters during parameter estimation.
There will also be a stochastic piece which varies from source to source.  This will
limit the precision at which $\chi_s$ can be measured, thus limiting the
resolution of dispersion-space density maps in the radial direction.
We note that one-point statistics from the same data set 
should provide empirical information about these properties independent
of their physical origin.

Detailed treatment of the above source modelling is beyond the scope of this
Letter. The magnitude of the contribution from the progenitor is, at present,
unknown. To crudely deal with it here, we use relatively large
bin widths of $\Delta\chi = 100\,{\rm Mpc}/h$
and exclude the $\chi = \chi'$ bin pairs.
Because the local term drops rapidly with radial separations, this effectively 
discards the contribution from the local term which is most sensitive to having
precise radial distances.

In first measurements of the angular power spectra, uncertainty will be
dominated by the finite number of observed events, or shot noise.
The noise power spectrum on shells at a distance $\chi_i$ and with width
$\Delta \chi$ is
\begin{equation}
    C^N_{\ell, ij} = \frac{\delta_{ij}}{\bar n_f(\chi_i)\chi_i^2\Delta\chi}.
\end{equation}
Here, to indicate that we are now working with quantities binned in the radial
direction, we have switched to $C_{\ell,ij}$ instead of
$C_\ell(\chi, \chi')$. The uncertainty on the angular
cross-power spectrum is then
\begin{align}
    (\Delta{C^{ss}_{\ell,ij}})^2 = 
    {\frac{1}{\ell(\ell + 1)\Delta \ell f_{\rm sky}}}
        \left[C^N_{\ell,ii} C^N_{\ell,jj} +  (C^N_{\ell,ij})^2\right],
\end{align}
where the second term in brackets only contributes for $i = j$.

\begin{figure}
    \includegraphics[scale=0.42]{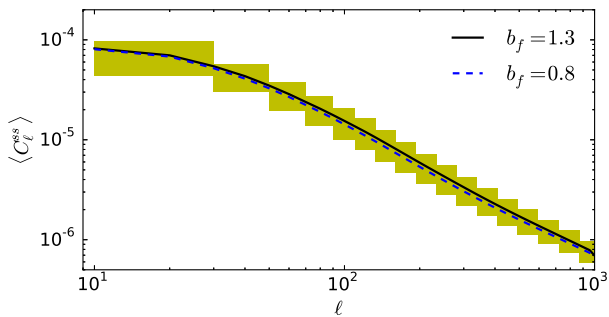}
    \caption{
        \label{f:sensitivity}
        Sensitivity of a survey with $10\,000$ dispersion measures distributed
        uniformly over half of the sky.  Plotted is the cross-correlation
        power spectrum weighted averaged over all pairs of radial bins. Weights
        are chosen to maximize signal to noise at $\ell=100$ for the $b_f=1.3$
        case.
    }
\end{figure}

We consider the sensitivity of a survey with a total of $10\,000$ dispersion
measures from FRB events distributed uniformly over half of the sky.
To give an idea of the overall sensitivity of the survey, we collapse
$C^{ss}_{\ell,ij}$ to a single function of $\ell$ by taking a weighted average
of all pairs of radial bins over the
range $500\,{\rm Mpc}/h$ to $3500\,{\rm Mpc}/h$.
For weights we use the signal over noise squared
at $\ell=100$, which maximizes the signal to noise at that $\ell$ and
is near optimal at other multipoles.
\blue{
To determine if such a survey would be
able to distinguish between models of FRB clustering we plot the same quantity
for $b_f=0.8$.
}
The result is shown
in Fig.~\ref{f:sensitivity}.

\section{Discussion and conclusions}

We have proposed using cosmological FRB sources as standard pings to trace the 3D
large-scale structure of the Universe using dispersion distance. It can be seen in
Fig.~\ref{f:terms} that the signal is dominated by inhomogeneities in the
electron density along the different lines of sight,
inducing misestimates of radial
distance, and thus apparent clustering.
\blue{
Because it is the line-of-sight integrated structure that correlates, the
signal is fairly insensitive to radial distance and the radial separation of
slices, with these dependencies being dominated by the coefficient $A(\chi)$.
}
These dispersion-space distortions could
be used
to study the distribution of free electrons in the Universe on large scales.
There have been other proposals to use FRBs for cosmology 
\citep{2014ApJ...780L..33M,2014MNRAS.443L..11D,
2014ApJ...783L..35D,2014PhRvD..89j7303Z,2014ApJ...788..189G}; however these
schemes have either not been three dimensional
or required externally measured redshifts
for the sources.

We have
shown in Fig.~\ref{f:sensitivity}
that a survey detecting $10\,000$ fast radio bursts of cosmological
origin could detect the
clustering signal.
This is in rough
approximation to what might be achieved in a reasonable amount of time by the
Canadian Hydrogen Intensity
Mapping Experiment (CHIME)\footnote{\url{http://chime.phas.ubc.ca/}}, the
Square Kilometre Array\footnote{\url{http://www.skatelescope.org/}},
or other upcoming wide-field telescopes.
\blue{Such a survey would be primarily sensitive to the integral 
term and thus rather insensitive to the details of FRB clustering models.}
Looking forward, dispersion-space
clustering could become a precision probe of large-scale structure should
observational factors turn out to be favorable compared to spectroscopic
surveys. The baryon acoustic oscillation feature can be clearly seen in
Fig.~\ref{f:terms} in the cross term and both
local-term curves. It may also be possible to extract the baryon acoustic
oscillation feature
from the integral
term using tomography.
This, however, requires that the statistics of the contribution to dispersion from
the source's environment and host halo to be well understood or small compared to the
contribution from the line of sight.

Finally, we note that alone---or in combination with
redshift data on a subset of the population, independent redshift surveys,
or gravitational lensing surveys---dispersion-space surveys could yield very
powerful probes of cosmology. Cosmological dispersion-space data could open a
new window into the structure of the Universe.

\begin{acknowledgments}

We thank Ue-Li Pen and Mathew McQuinn the for helpful discussions.
K.W.\,M is supported by the Canadian Institute for Advanced Research, Global
Scholars Program.  The research of K.\,S. is supported in part by a National
Science and Engineering Research Council (NSERC) of Canada Discovery Grant.

\end{acknowledgments}

\bibliography{refs}

\end{document}

%% file: AAS_macros.tex
\let\jnl@style=\rm
\def\ref@jnl#1{{\jnl@style#1}}

\def\aj{\ref@jnl{AJ}}                   
\def\araa{\ref@jnl{ARA\&A}}             
\def\apj{\ref@jnl{ApJ}}                 
\def\apjl{\ref@jnl{ApJ}}                
\def\apjs{\ref@jnl{ApJS}}               
\def\ao{\ref@jnl{Appl.~Opt.}}           
\def\apss{\ref@jnl{Ap\&SS}}             
\def\aap{\ref@jnl{A\&A}}                
\def\aapr{\ref@jnl{A\&A~Rev.}}          
\def\aaps{\ref@jnl{A\&AS}}              
\def\azh{\ref@jnl{AZh}}                 
\def\baas{\ref@jnl{BAAS}}               
\def\jrasc{\ref@jnl{JRASC}}             
\def\memras{\ref@jnl{MmRAS}}            
\def\mnras{\ref@jnl{MNRAS}}             
\def\pra{\ref@jnl{Phys.~Rev.~A}}        
\def\prb{\ref@jnl{Phys.~Rev.~B}}        
\def\prc{\ref@jnl{Phys.~Rev.~C}}        
\def\prd{\ref@jnl{Phys.~Rev.~D}}        
\def\pre{\ref@jnl{Phys.~Rev.~E}}        
\def\prl{\ref@jnl{Phys.~Rev.~Lett.}}    
\def\pasp{\ref@jnl{PASP}}               
\def\pasj{\ref@jnl{PASJ}}               
\def\qjras{\ref@jnl{QJRAS}}             
\def\skytel{\ref@jnl{S\&T}}             
\def\solphys{\ref@jnl{Sol.~Phys.}}      
\def\sovast{\ref@jnl{Soviet~Ast.}}      
\def\ssr{\ref@jnl{Space~Sci.~Rev.}}     
\def\zap{\ref@jnl{ZAp}}                 
\def\nat{\ref@jnl{Nature}}              
\def\iaucirc{\ref@jnl{IAU~Circ.}}       
\def\aplett{\ref@jnl{Astrophys.~Lett.}} 
\def\apspr{\ref@jnl{Astrophys.~Space~Phys.~Res.}}
\def\bain{\ref@jnl{Bull.~Astron.~Inst.~Netherlands}} 
\def\fcp{\ref@jnl{Fund.~Cosmic~Phys.}}  
\def\gca{\ref@jnl{Geochim.~Cosmochim.~Acta}}   
\def\grl{\ref@jnl{Geophys.~Res.~Lett.}} 
\def\jcp{\ref@jnl{J.~Chem.~Phys.}}      
\def\jgr{\ref@jnl{J.~Geophys.~Res.}}    
\def\jqsrt{\ref@jnl{J.~Quant.~Spec.~Radiat.~Transf.}}
\def\memsai{\ref@jnl{Mem.~Soc.~Astron.~Italiana}}
\def\nphysa{\ref@jnl{Nucl.~Phys.~A}}   
\def\physrep{\ref@jnl{Phys.~Rep.}}   
\def\physscr{\ref@jnl{Phys.~Scr}}   
\def\planss{\ref@jnl{Planet.~Space~Sci.}}   
\def\procspie{\ref@jnl{Proc.~SPIE}}   